\documentclass[12pt,reqno]{article}
\usepackage{amsmath}
\usepackage{amsthm}
\usepackage{amsfonts}
\usepackage{amssymb}
\usepackage{epic}
\usepackage{eepic}
\usepackage{graphics}
\newcommand{\Y}{\Upsilon}
\newcommand{\la}{\lambda}
\newcommand{\ra}{\rightarrow}
\newcommand{\bz}{\bar{z}}
\def\ta{\tilde\a}
\def\t{\tau} 
 
\def\c{\gamma} 
\def\pl{\partial} 
 
\def\QR{\mathbb{R}} 
\def\QC{\mathbb{C}} 
\def\a{\alpha}
\def\b{\beta} 
\def\c{\gamma} 
 
\def\e{\epsilon} 
 
\newcommand{\beq}{\begin{equation}}
\newcommand{\eeq}{\end{equation}}

\def\beas{\begin{eqnarray*}}
\def\eeas{\end{eqnarray*}}
\def\bea{\begin{eqnarray}}
\def\eea{\end{eqnarray}}
\def\a{\alpha} 
\def\b{\beta} 
 
\newcommand{\remlst}{\begin{list}
{(\arabic{num})}{\usecounter{num}\topsep0cm \itemsep0cm \parsep0cm}}
\title{\bf Rolling Tachyons from Liouville theory}  
\author{\\[5mm] Volker Schomerus \\[5mm]Service de Physique 
Th\'eorique, CEA Saclay,\\ F-91191 Gif-sur-Yvette, France\\[5mm] }
\date{June 3, 2003} 
\begin{document}
\begin{titlepage}      \maketitle       \thispagestyle{empty}

\vskip1cm
\begin{abstract} 
In this work we propose an exact solution of the 
$c=1$ Liouville model, i.e.\ of the world-sheet theory that 
describes the homogeneous decay of a closed string tachyon. 
Our expressions are obtained through careful extrapolation 
from the correlators of Liouville theory with $c \geq 25$.
In the $c=1$ limit, we find two different theories which 
differ by the signature of Liouville field. The Euclidean 
limit coincides with the interacting $c=1$ theory that was 
constructed by Runkel and Watts as a limit of unitary minimal 
models. The couplings for the Lorentzian limit are new. In 
contrast to the behavior at $c > 1$, amplitudes in both $c=1$ 
models are non-analytic in the momenta and consequently they 
are not related by Wick rotation.
\end{abstract} 

\vspace*{-15.9cm}
{\tt {SPhT-T03/074}}
\bigskip\vfill
\noindent
\phantom{wwwx}{\small e-mail:}{\small\tt
vschomer@spht.saclay.cea.fr} 

\end{titlepage} 

\baselineskip=19pt 
\setcounter{equation}{0} 
\section{Introduction}
\def\tr{{\rm tr}}

Over the last years we have made enormous progress in our 
understanding of the condensation processes that are triggered 
by closed and open string tachyons. Most of the results are 
based on the conjecture that the world-sheet models which 
realize the initial and final state of these condensation 
processes are related by renormalization group (RG) flows. 
This conjectured relation between space-time dynamics and 
world-sheet RG flows is still far from being understood, 
especially in the case of closed string tachyons (see e.g.\ 
\cite{Martinec:2002tz,Gutperle:2002ki} for a review of 
recent progress and references). Many of the open 
issues can only be addressed through the construction of 
exact time dependent backgrounds. For the bounce of an open 
string tachyon such investigations were initiated by A.\ Sen
\cite{Sen:2002nu,Sen:2002vv}. Studies of the closely related 
rolling tachyon solution mainly go back to the work of A.\ 
Strominger et al.\ \cite{Strominger:2002pc,Gutperle:2003xf,
Strominger:2003fn}. This line of research has continued to  
provide important new insights into various aspects of open 
string tachyon condensation (see e.g.\  \cite{Mukhopadhyay:2002en,
Okuda:2002yd,Chen:2002fp,Larsen:2002wc,Rey:2003xs,Lambert:2003zr}), 
in particular for 2D string theory where quantum 
corrections to tree level results can be incorporated due to 
the duality with matrix models \cite{McGreevy:2003kb,
Martinec:2003ka,Klebanov:2003km,McGreevy:2003ep}. 
\smallskip

In most studies of such exact decay backgrounds, attention was
restricted to special quantities in the world-sheet model, primarily 
to the boundary states, and very few attempts have been made to 
obtain the full solution, including the bulk and boundary 2- and 
3-point couplings. This is the stage on which the results of this 
note are set. Our aim here is to obtain the full exact solution of 
the rolling closed string tachyon background. While closed string 
tachyons have certainly been 
a central issue, in particular for bosonic string theory, their 
condensation processes are rather difficult to picture, mainly 
because of the drastic effects they have on space-time itself. 
It therefore seems that the study of open string tachyons is a 
much better starting point. On the other hand, exact solutions 
of the boundary conformal field theories that appear in the 
description of branes and open strings are often facilitated 
if an appropriate basis of bulk states is chosen, associated with 
a corresponding bulk interaction. This is our main motivation 
to study the bulk theory first and we will see below how our 
results indeed represent a crucial step toward the solution 
of the boundary problem. 
\smallskip

Before we state our results, let us recall that tachyon 
instabilities of a Lorentzian $D+1$-dimensional static string 
background are related to relevant world-sheet fields 
$\Phi(z,\bz)$ of a unitary conformal field theory with 
$D$-dimensional Euclidean target through 
$$ \delta S \ = \ \int_\Sigma dzd\bz \, e^{i E_\Phi X_0} \, 
   \Phi(z,\bz)  \ \ \ \ \mbox{ where } \ \ \ \ 
   E^2_\Phi \ = \ \Delta_\Phi-2 
 \ \ . $$ 
Here, $X_0$ denotes the time-like free bosonic field that 
represents the time coordinate and $\Delta_\Phi = h_\Phi + 
\bar h_\Phi$ is the conformal weight of $\Phi$. For relevant
or marginally relevant fields, $\Delta_\Phi \leq 2$ implies
a purely imaginary $E_\Phi$ so that $\delta S$ represents an 
admissible perturbation of the original static background. 
The simplest example arises when $\Phi(z,\bz)$ is the identity 
field in which case $\delta S$ does not couple the unitary 
spatial conformal field theory with the time component. Hence, 
we can focus on the perturbation of the time-like free boson 
$X_0$ with central charge $c=1$. The whole setup resembles
very much the problem of constructing Liouville theory with 
$c=1$, only that now the signature of the Liouville direction 
$X$ differs from the usual situation. As a consequence, the 
solution of the rolling tachyon background requires to obtain 
world-sheet correlators for the fields $\exp( 2 \a X) \sim 
\exp (-2 i\a X_0)$ with real rather than imaginary parameter 
$\a$.
\smallskip 

What gives us hope to cope with this problem is the fact that Liouville 
has been solved for $c \geq 25$ in several steps throughout the 
last years \cite{Dorn:1994xn,Zamolodchikov:1996aa,Ponsot:1999uf,
Teschner:2001rv,Teschner:2003en}.  
Moreover, the solution is analytic, both in its dependence on the
central charge and on the labels $\a$. The latter fact implies that
all couplings can be continued to imaginary $\a$ and hence it seems
irrelevant that the Liouville direction is usually taken to be 
space-like. The analytic dependence on the central charge, on the 
other hand, may be used to continue the exact solution down to 
$c > 1$. But unfortunately, at the point $c=1$, poles and zeros
of the amplitudes move on top of each other, a behavior which has  
led to the widespread believe that the $c=1$ limit is not 
well-defined. As we shall see below, this is not the case! 
In fact, theories
with $c \leq 1$ can be extracted from a rigorous limit of the 
$c \geq 25$ solution, but the resulting models possess 
several unexpected and interesting features. As we shall argue, 
for $c \leq 1$ they do no longer depend smoothly on the central 
charge $c$. More importantly, their couplings are not analytic 
in the momenta $\a$. This implies in particular that Euclidean 
(E) and Lorentzian (L) theories cannot be related by Wick rotation, 
i.e.\ by a continuation in the parameter $\a$. Similar issues 
have also been discussed recently in \cite{Gaiotto:2003rm}, but 
in our case there exists a quite natural prescription to resolve 
the problem: If we want to go from the Euclidean $c = 1$ theory to 
the Lorentzian rolling tachyon solution, we first move the central 
charge back into the regime $c \geq 1$, continue analytically to 
real $\a$ and take then take the limit $c \ra 1$. In this way we 
shall find two interacting $c=1$ theories, one for imaginary and 
the other for real values of the parameter $\a$.  
\smallskip 

\def\X{{\rm S}} \def\L{{\rm L}} \def\E{{\rm E}} 
The 3-point couplings $C(\a_1,\a_2,\a_3)$ we propose for these two 
theories contain two different factors. One of them is analytic in 
the couplings $\a_j$ while the other is built from step functions. 
More precisely, we shall find that 
\begin{eqnarray} & & \hspace*{-2cm}  \nonumber 
 \langle \, e^{2\a_1 X(z_1,\bz_1)} \, e^{2 \a_2 X(z_2,\bz_2)} \, 
   e^{2\a_3 X(z_3,\bz_3)}\rangle \ = \ \frac{ C^\X_{c=1}(\a_1,\a_2,\a_3)} 
   {|z_{12}|^{2h_{12}} |z_{23}|^{2h_{23}} |z_{13}|^{2h_{13}}}  \\[2mm] 
 \mbox{ where } \ \  z_{ij} & = & z_i - z_j \ \ \ , \ \ \ 
   h_{ij} \ = \ \a_i^2 +\a^2_j -\a^2_k \ \ \mbox{ for } \ \ 
  k \ \neq \ i,j\ \ \mbox{and} \nonumber  \\[6mm]       
 C^\X_{c=1}(\a_1,\a_2,\a_3) & = & (\pi \mu_{\rm ren})^{i\sum_j\a_j} \, 
 \, e^{- 2 \pi \ta} \    
 P^\X_{c=1}(\a_1,\a_2,\a_3) \times \label{Cint} \\[2mm] 
 & & \times  \exp \int_0^\infty \frac{d\tau}{\tau}
 \, \frac{1}{\sinh^2 \frac{\tau}{2}} 
 \left[ \sin^2 \ta \tau + \sum_{j=1}^3 \, (\sin^2 \ta_j \tau 
   - \sin^2 \a_j \tau)\right] \ \ .\nonumber 
\end{eqnarray} 
Here $\mu_{\rm ren}$ was introduced as in \cite{Strominger:2003fn} 
and the quantities $\ta, 
\ta_j$ are certain linear combinations of the parameters $\a_j$ to be 
defined in eqs.\ (\ref{tdef1}),(\ref{tdef2}). The superscript $\X$ 
stands for $\X = \L,\E$. As we explained above, the Lorentzian theory 
$\X=\L$ requires $\a_j \in \QR$ while in the Euclidean case $\X=\E$ 
we must choose $\a_j = i \eta_j$ with $\eta_j \in \QR$. The factor 
$P^\X$ is finally given by 
$$ P^\X_{c=1}(\a_1,\a_2,\a_3) \ = \ e^{4\pi \ta} 
\ \left( 1 + \Theta^\X(\ta) \, \prod_{j=1}^{3} \, \Theta^\X(\ta_j)
 \right)\ \ 
$$   
where $\Theta^\L(\a) = \Theta(\a)$ and $\Theta^\E(\a) = 
\theta(-i\alpha)$ are step functions $\Theta$ and $\theta$ 
on the real line, see eqs.\ (\ref{Theta}) and (\ref{theta}). 
In writing this expression for $P^\X$ we have dropped some 
constant pre-factors from the formulas below. Such factors can 
always be absorbed in the normalization of the fields. In the 
Euclidean case, our formulas reproduce the couplings of an 
interacting $c=1$ model that was constructed by Runkel and Watts 
in \cite{Runkel:2001ng}. The Lorentzian theory was previously 
considered by Strominger and Takayanagi \cite{Strominger:2003fn}. 
A comparison with formula (4.17) of the latter paper shows that 
the two proposals differ mainly by the factor $P^\L$. \footnote{
It seems that the factor $\exp I$ (see eqs.\ (4.7),(4.8) of
\cite{Strominger:2003fn}) has been accidentally left out of the  
final proposal (4.17) of \cite{Strominger:2003fn} for 
the 3-point coupling.}   
\medskip 
  
We shall begin our analysis with a short section on the minisuperspace
model in which we review some of the results from \cite{Strominger:2002pc} 
before 
computing a minisuperspace analogue of the 3-point coupling. The 
latter turns out to share some important features with the exact 
solution. In the third section we recall all necessary facts 
about the solution of Liouville theory with $c \geq 25$. The 
3-point couplings will be rewritten in section 4 in a way that is 
well adapted to taking the central charge down to values $c \leq 1$. 
This enables us to make a precise proposal for the 3-point functions
of $c \leq 1$ models. An explicit computation of our expressions 
boils down to the evaluation of a certain ratio of Jacobi 
$\vartheta$-functions $\vartheta_1(x,\tau)$ at points $q = \exp 
\pi i \tau$ on the boundary $|q|=1$ of the unit disc. For the
$c=1$ theory, we perform these calculations in the last section. 
This leads to the formulas we have spelled out above.

\section{The minisuperspace analysis}
\setcounter{equation}{0} 
\def\w{\omega} 
\def\tw{\tilde \w}

Before we start to descend into our analysis of the exact conformal 
field theories we would like to study a much simpler toy model that 
we will later recover as a `minisuperspace limit' of the full solution. 
The analysis of the toy model was initiated by Strominger in 
\cite{Strominger:2002pc}. Here, we shall review some elements of the 
model and then provide new expressions for an analogue of its 3-point 
function.   
\medskip 

To gain some insights into the $c=1$ Liouville theory, Strominger 
suggested to study solutions $\phi$ of the following Schroedinger 
equation with $\la >0$,  
\begin{equation}
 H_0\, \phi \ :=\ \left(\frac{\pl^2}{\pl x^2_0} + \lambda \, 
     e^{2 x_0} \right) \, \phi(x_0) \ = \ - 4 \omega^2 \phi(x_0)\ \ . 
\end{equation}
The unusual sign in front of the derivative is associated with the 
fact that $x_0$ is thought of as a time-coordinate. As a consequence, 
the second term appears as if there was a potential that is unbounded
from below. This certainly is reflected in the structure of the 
solutions which are given by arbitrary linear combinations of the 
function 
\begin{equation} \label{solin} 
 \phi^{\rm in}_\omega(x_0) \ = \ (\lambda/4)^{i\w} 
   \, \Gamma(1-2 i\omega) \, 
   J_{-2i\omega}(\sqrt{\lambda}e^{x_0}) \ \ . 
\end{equation}  
and its complex conjugate. Here, $J_\nu$ is a Bessel function of the 
first kind. We have normalized the wave function such that in the far past 
$$ \phi^{\rm in}_\omega(x_0) \ \sim \ e^{-2 i \omega x_0} \ \ \ 
   \mbox{ for } \ \ \ x_0 \ra -\infty \ \ .  
$$ 
In the far future, on the other hand, the solution $\phi^{\rm in}_\omega$
takes the form
$$ \phi^{\rm in}_\omega(x_0) \ \stackrel{x_0 \ra \infty}{\sim} \  
   (\lambda/4)^{i \omega -\frac{1}{4}} \, 
   \Gamma(1-2i\omega) \, \frac{e^{-x_0/2}}{\sqrt{4\pi}}\,    
   \left(e^{i \sqrt \lambda e^{x_0} - \frac{\pi}{4}(4\omega + i)}
   +  e^{-i \sqrt \lambda e^{x_0} + \frac{\pi}{4}(4\omega + i)}\right) \ \ .
$$
Hence, positive and negative frequencies appear as we proceed in time. 
This can be considered as a signal for pair production 
\cite{Strominger:2002pc} and we will explain how to read off the 
production rate below. 
\smallskip 

In \cite{Gutperle:2003xf} is was argued that the wave functions which 
correspond to fields in Liouville theory are special linear combinations
of the wave function (\ref{solin}) and its complex conjugate. The reasoning 
is as follows: Liouville theory possesses an exponential potential with 
positive coefficients. Hence, among the two linear independent solutions 
there is one that decays exponentially while the other diverges at the 
same rate. The latter is clearly unphysical and so we end up with a 
single physical solution which, after Wick rotation, contains only 
positive frequencies. It corresponds to the function   
$$ \phi_\omega(x_0) \ = \ (\lambda/4)^{i \omega} \,  
   \Gamma(1-2i\omega)\, \left(\,  
    J_{-2i\omega}(\sqrt{\lambda} e^{x_0})
   -  e^{- 2 \pi \omega}\,  
   J_{2i\omega}(\sqrt{\lambda} e^{x_0})\right) \ \  
$$ 
with $\omega \geq 0$. An analogue of the 2-point function for these wave 
functions and its relation to the pair creation rate was discussed in 
\cite{Strominger:2002pc}. 
\medskip

Here we are mainly interested in the 3-point function, since in the full 
conformal field theory this quantity encodes all the information about 
the exact solution. Its counterpart in the minisuperspace model can be 
evaluated through the following integral over a product of Bessel 
functions,   
\begin{eqnarray} 
 \langle \w_1| e_{\w_2} |\w_3\rangle & := & 
    \int_{-\infty}^\infty dx_0\ \phi_{\w_1}(x_0) \, e^{-2 i \w_2 x_0}\,  
    \phi_{\w_3}(x_0)  \ = \ 
    (\lambda/4)^{2 i \tw} \, 
   P_0(\w_j) \, e^{Q_0(\w_j)} \nonumber \\[2mm] 
  \mbox{where} \ \ \ \ \ \ & &    
    \exp Q_0(\w_1, \w_2, \w_3) \ = \ \frac{1}{\Gamma(1+2 i\tw)} 
     \ \prod_{j=1}^3 
   \frac{\Gamma(1+(-1)^j 2 i\w_j)}{\Gamma(1-(-1)^j 2 i\tw_j)}\ \ , 
    \label{mss3}\\[2mm]
     P_0(\w_1, \w_2, \w_3) & = &  
  \frac{4\pi i}{\sinh 2 \pi\tw} + \sum_{j=1}^3 \, 
   \frac{4\pi i \, e^{(-1)^j 2 \pi\tw_j-2 \pi \tw}}{\sinh2 \pi\tw_j} 
   \ = \ - \frac{2\pi i\, e^{-2 \pi \tw}}{\sinh 2 \pi \tw} \ \prod_{j=1}^3 
   \, \frac{\sinh 2 \pi \w_j}{\sinh 2 \pi \tw_j} \ \ . \nonumber
\end{eqnarray}
The formula that was used to compute the integral can be found in standard 
mathematical tables. To write the answer more compactly, we have assigned  
the 4-tuple $\tw,\tw_j$ to the triple $\w_j, j = 1,2,3,$ by the simple 
prescription
\begin{eqnarray} \label{tdef1} 
 \tw \ = \ \frac12(\w_1+\w_2+\w_3) & \ \ \ \ \ \  ,  \ \ \ \ \ \ & 
 \tw_1 \ = \ \frac12(\w_2+\w_3-\w_1) \\[2mm] \label{tdef2} 
 \tw_2 \ = \ \frac12(\w_3+\w_1-\w_2) & , & 
 \tw_3 \ = \ \frac12(\w_1+\w_2-\w_3) \ \ .  
\end{eqnarray} 
Splitting the result of the integration in eq.\ (\ref{mss3}) into the 
two factors $P_0$ and $\exp Q_0$ may seem a bit artificial at this point. 
But we shall see below that the split arises quite naturally when we 
recover the same expression through the minisuperspace limit of the exact 
theory. 
\medskip  

From the 3-point function it is not hard to extract the 2-point function 
of our toy model in the limit $\omega_2 = \varepsilon \ra 0$. Note that 
the factor $P$ has poles at $\omega_1 - \omega_3 \pm \varepsilon = 0$ 
which, after taking the limit, produce the $\delta(\omega_1-\omega_3)$ 
that we expect to find. More precisely, for $\omega_1 , \omega_3 \geq 0$
we obtain  
$$ \lim_{\omega_2 \ra 0} \, C_0(\w_1,\w_2,\w_3) \ \sim \ 
  \delta(\w_1-\w_3) \,  e^{-2 \pi \w_1}\, (\lambda/4)^{2 i \w_1} \ 
 \frac{\Gamma(1- 2 i\omega_1)}{\Gamma(1+2 i\omega_1)}\ \ . 
$$  
The modulus of the quantity that multiplies the $\delta$-function was 
interpreted in \cite{Strominger:2002pc} as the pair production rate of 
this toy model.  

\section{Liouville theory with $c\geq 25$}
\setcounter{equation}{0}

The aim of this section is to review the solution of Liouville theory 
with $c \geq 25$. This solution was first proposed several years ago by 
H.\ Dorn and H.J.\ Otto \cite{Dorn:1994xn} and by A.\ and Al.\ Zamolodchikov 
\cite{Zamolodchikov:1996aa} after M.\ Goulian and M.\ Li 
\cite{Goulian:1991qr} had taken some intermediate step. Crossing symmetry 
of the conjectured 3-point function was then checked analytically in two 
steps by Ponsot and Teschner \cite{Ponsot:1999uf} and by Teschner 
\cite{Teschner:2001rv,Teschner:2003en}.   
\medskip

As in any bulk conformal field theory, the exact solution of Liouville 
theory is entirely determined by the structure constants of the 3-point 
functions for the (normalizable) primary fields 
\begin{equation}\label{flds} 
 \Phi_\a(z,\bz) \ \sim \ e^{2 \a X(z,\bz)} 
\ \ \ \mbox{ with } \ \ \a \ = \ \frac{Q_b}{2} + i p
 \ \ \ , \ \ \ p\ \geq \ 0 \ \  \end{equation}
where $Q_b = b + b^{-1}, b \in \QR$. These fields are primaries 
with conformal weight 
$h_\a = \bar h_\a= \a (Q_b-\a)$ under the action of the two Virasoro 
algebras whose central charge is $c = 1 + 6 Q^2_b$. The couplings of 
three such fields are given by the following expression 
\cite{Dorn:1994xn,Zamolodchikov:1996aa}.  
\begin{equation}\label{Cc>} 
C(\a_1,\a_2,\a_3) \ := \ \left[\pi \mu \gamma(b^2) b^{2-2b^2}
\right]^{(Q_b-2 \ta)/b} \ \frac{\Y'(0)}{\Y(2\ta-Q_b)} \ \prod_{j=1}^3 
  \, \frac{\Y(2\a_j)}{\Y(2\ta_j)}
\end{equation}  
where $\ta$ and $\ta_j$ are the linear combinations of $\a_j$ which are
introduced just as in eqs.\ (\ref{tdef1}),(\ref{tdef2}) of the previous 
section,  $\gamma(y) = \Gamma(y)/\Gamma(1-y)$ is a quotient of ordinary 
$\Gamma$-functions and the function $\Y= \Y_b$ is defined in terms of 
Barnes' double $\Gamma$-function $\Gamma_2$ (see appendix 1) by 
\begin{equation} \label{ZY} 
 \Y_b (\a) \ := \ \Gamma_2(\a|b,b^{-1})^{-1}\, 
                    \Gamma_2(Q_b-\a|b,b^{-1})^{-1} \ \ . 
\end{equation}  
The properties of the double $\Gamma$-function which we spell out in 
appendix 1 imply that $\Y$ possesses the following integral representation
\begin{equation}\label{ZYpsilon} 
\ln \Y_b(y) \ = \ -2c_b + \int_0^\infty \frac{dt}{t} 
   \left[ \left(\frac{Q_b}{2}-y\right)^2
     e^{-t} - \frac{\sinh^2\left(\frac{Q_b}{2} -y\right) \frac{t}{2}}
        {\sinh\frac{bt}{2}\, \sinh \frac{t}{2b}}\right]\ \ .  
\end{equation}
The constant $c_b$ is defined in appendix 1, but its value will not be 
relevant below. Moreover, we deduce from the two shift properties 
(\ref{BGshift}) of the double $\Gamma$-function that 
\begin{equation}\label{ZYshift} 
   \Y_b(y+b) \ = \ \gamma(by) \, b^{1-2by}\, \Y_b(y) \ \ , \ \ 
   \Y_b(y+b^{-1}) \ = \  \gamma(b^{-1}y) \, b^{-1+2b^{-1}y}\, \Y_b(y)
\end{equation}
Note that the second equation can be 
obtained from the first with the help of the self-duality property $\Y_b(y) = 
\Y_{b^{-1}}(y)$. We would like to stress that poles in Barnes double 
$\Gamma$-function give rise to zeros of the function $\Y$. The latter 
induce poles in the dependence of the coefficients (\ref{Cc>}) on the 
variables $\ta_j$ and $\ta$. This fact will become important below. 
\medskip

There are two further remarks that we would like to add. To begin with, we 
observe that the couplings (\ref{Cc>}) obey the following shift equation
\begin{equation}\label{Cshift} 
    \frac{C(\a_1 + b,\a_2,\a_3)}{C(\a_1,\a_2,\a_3)} \ = \ 
   -\frac{\c(-b^2)}{\pi \mu} \, \frac{\c(b(2\a_1+b))\c(b2\a_1)}
    {\c(b(2\ta - Q_b))} \ \prod_{j=1}^3 \, \frac{1}{\c(b2\ta_j)} \ \ 
\end{equation}
and a dual equation with $b$ being replaced by $b^{-1}$. This equation can be 
derived from the factorization constraint for the 4-point function of the 
three fields $\Phi_{\a_j}$ with the fundamental Liouville field $\Phi = 
\Phi_{b/2}$. The derivation exploits the equation of motion of the Liouville 
field, see \cite{Teschner:1995yf} for details. What we want to point out is 
that the two shift 
equation (\ref{Cshift}) and its dual counterpart possess a unique smooth 
(in $b$) family of solutions. Hence, the equations of motion along with the 
self-duality of Liouville theory fix the expression (\ref{Cc>}) for the 
3-point function, given that the model exists for all $c \geq 25$ and 
that it depends smoothly on the central charge. 
\smallskip

Let us finally stress that the expression (\ref{Cc>}) can be evaluated for 
values $\a_j$ other than those given in eq.\ (\ref{flds}). Even though the
corresponding objects do not correspond to normalizable states of the model, 
they may be considered as well defined but non-normalizable fields. It is 
tempting to identify the identity field with limit $\lim_{\a \ra 0} 
\Phi_{\a}$. This identification can indeed be confirmed by computing 
the corresponding limit of the coefficients $C$ which is given by 
\begin{eqnarray} \label{2ptcg}   
\lim_{\a_2 \ra 0} C(\a_1,\a_2,\a_3) & = & 2\pi \, \delta(\a_1 +\a_3 - Q_b) 
  \, + \, R(\a_1) \, \delta(\a_1-\a_3)  \\[2mm] \mbox{where} \ \ \ \ \ \ 
 R(\a) & = &  \left(\pi \mu \c(b^2)\right)^{(Q_b-2\a)/b}\ 
  \frac{b^{-2} \c(2b\a - b^2)}{\c(2-2b^{-1}\a + b^{-2})}\ \  
\nonumber 
\end{eqnarray}
for all $\a_{1,3} = Q_b/2 + i p_{1,3}$ with $p_i \in \QR$. If we restrict 
to the region where $p_1,p_3 \geq 0$, only the second term containing the 
{\it reflection amplitude} $R(\a)$ survives. The $\delta$-functions again 
emerges from the singularities of $C$, just as in the minisuperspace
example. We would like to stress that the reflection amplitude can also 
be obtained directly from the 3-point coupling without ever performing 
a limit $\a_2 \ra 0$. In fact, in Liouville theory the fields with label 
$\a$ and $Q_b-\a$ differ only by a relative complex factor (see e.g.\ 
\cite{Teschner:2001rv} for a more detailed discussion), namely by the 
factor $R(\a)$, i.e.\ 
$$ C(\a_1,\a_2,\a_3) \ = \ R(\a_1) \ C(Q-a_1,\a_2,\a_3) \ \ . $$ 
This remark shall become quite relevant below since it allows to recover 
the reflection amplitude without constructing the identity field first. 
Note that for the latter we had to continue correlation functions from 
the line $Q_b/2 + i \QR$ into the complex plane.  

\section{On exact solutions for $c\leq 1$} 
\setcounter{equation}{0}

We are now prepared to advance toward the solution of Liouville theory 
for $c \leq 1$. Our strategy is to continue the theory analytically from 
the regime $c \geq 25$ down to smaller central charges. This is achieved 
through analytic continuation of the parameter $b$ into the complex plane. 
At first,  such a 
continuation seems quite straightforward: as we move $b$ along the unit 
circle, the central charge
drops below $c=25$ while the expression for the 3-point function (\ref{Cc>})
remains well defined. In fact, the construction of Barnes double 
$\Gamma$-function $\Gamma_2$ is not problematic unless ${\rm Re}(b) = 0$. 
Unfortunately, the values $c \leq 1$ of the central charge that we are 
interested in correspond precisely to values of $b$ along the imaginary 
axis for which Barnes double $\Gamma$-function is not defined. Note, 
however, that the 2-point function (\ref{2ptcg}), the shift equation
(\ref{Cshift}) and its dual only involve ordinary $\Gamma$-functions 
and hence they possess a smooth continuation into the regime with 
$c\leq 1$.%
\smallskip

The last remark seems to suggest that we should simply look for a new
solution of the shift equations (\ref{Cshift}) that we can use when 
$c \leq 1$. It turns out that such a solution is rather easy to write
down. To this end, we introduce the following new function 
\begin{equation} \label{Y}   
  Y_\b(\a) \ := \ \Gamma_2(\b + i \alpha|\b,\b^{-1})\, 
                 \Gamma_2(\b^{-1}-i\a|\b,\b^{-1}) \ \ . 
\end{equation} 
This is defined for all $\b = -ib$  with ${\rm Re} (\b) \neq 0$. Again, 
using properties of Barnes double $\Gamma$-function we can deduce some
important properties of the function $Y$. In particular, it possesses 
the following integral representation 
\begin{equation}  \label{Yint} 
 \ln Y_\b(\a) \ = \ 2 c_\b + \int_0^\infty  \frac{d\t}{\t}
              \left[ e^{-\t}\left(\frac{Q_b}{2}-\a\right)^2   -
               \frac{ \sin^2\left(\frac{Q_b}{2}-\a\right)\frac{\t}{2}}
               {\sinh\frac{\b \t}{2} \sinh\frac{\t}{2\b}}    
               \right] 
\end{equation} 
with a constant $c_\b$ that is defined in appendix 1.  
Similarly, we obtain shift properties of our new function $Y$ from 
the behavior of Barnes double $\Gamma$-functions 
\begin{eqnarray}  \label{Yshift1} 
Y(\a+i\b) & = & \b^{1-2i\a\b}\, \gamma (i\a\b)\, Y(\a) \ \ , \\[2mm]
Y(\a-i\b^{-1}) & = & \b^{-1-2i\a\b^{-1}}\, \gamma (-i\a/\b)\, Y(\a)\ \ . 
\label{Yshift2} \end{eqnarray} 
Observe in passing that the second equation is not obtained from the 
corresponding shift equation (\ref{ZYshift}) by the formal substitution 
$b = i \b$. While this difference will become important below it is not 
relevant for our construction of a solution to eqs. (\ref{Cshift}) and 
its dual. In fact, it is easy to obtain such a solution through the 
following combination of $Y$-functions,  
\begin{equation} \label{Q}
\exp Q(\a_1,\a_2,\a_3)  \ := \  \frac{Y(0)}{Y(2\ta - Q)} 
\ \prod_{j=1}^3 \,  \frac{Y(2\a_j)}{Y(2\ta_j)}\ \ .  
\end{equation}  
Even though this is the unique smooth family of solutions to eq.\ 
(\ref{Cshift}) and its dual, we refrained from denoting it by $C$.
Actually, we shall argue momentarily that the expression (\ref{Q}) 
cannot give the exact 3-point function. This then allows us to draw 
our first interesting conclusion: In the regime $c \leq 1$, Liouville 
theory does not depend smoothly on the central charge. Such a
behavior is in sharp contrast to the properties of Liouville 
theory for $c > 1$.  
\smallskip 

There are two simple arguments that can be raised against an identification 
of the expression $\exp Q$ with the 3-point couplings. First of all, we 
want to stress that our function $Y$ is defined as a product of Barnes 
$\Gamma$-functions $\Gamma_2$ and not through $1/\Gamma_2$ as in 
the construction 
(\ref{ZY}) of the function $\Y$. Hence, the function $Y$ has a discrete set
of poles but no zeros. This in turn implies that the quantity $\exp Q$ 
has only zeros in the arguments $\ta$. But as we pointed out before, 
poles in $\ta$ are needed in order to recover the $\delta$-function in 
the 2-point function upon taking $\a_2$ to zero. Consequently, such a 
limit $\a_2 \ra 0$ of $\exp Q$ cannot give a sensible 2-point function. 
\smallskip 

Our second argument relies on a comparison with the toy model in section 2.  
To this end, we choose the parameters $\a_j$ to be of the form 
\begin{equation} \label{aom}  
 \a_1 \ = \ \frac{Q_b}{2}  + \omega_1 \beta \ \ , \ \ \a_3 \ = \ 
   \frac{Q_b}{2} + \omega_3 \beta   \ \ , \ \ \a_2 \ = \ \omega_2 
\beta \ \ . 
\end{equation} 
We can then take the limit $\beta \ra 0$ of our expression for $\exp Q$ 
using the following asymptotic behavior of our function $Y$,  
$$ Y_\b(\b y) \ \sim\  Y_0 \b^{iy} \, \Gamma(1+iy) + \dots \ \ , $$
along with the simple property $Y_\b(Q_b-\a) = Y_\b(\a)$. For the limit 
we then find that  
$$  
\lim_{\b \ra 0} e^{Q(\a_1,\a_2,\a_3)} \ = \  
\frac{1}{\Gamma(1+2 i\tw)} \ \prod_{j=1}^3 
  \frac{\Gamma(1+(-1)^j 2 i\w_j)}{\Gamma(1-(-1)^j 2 i\tw_j)} \ = \ 
  e^{Q_0(\omega_1,\omega_2,\omega_3)} \ \ . 
$$ 
Hence, we are missing entirely the factor $P_0$ that was an important 
piece of 3-point function in our toy model. Recall that the poles in 
the factor $P_0$ were crucial in recovering the 2-point function from 
the 3-point coupling $C_0$. 
\bigskip

In order to come up with a promising proposal for the 3-point function 
we will now try to understand how the function $\Y_b$ and our new 
function $Y_\b$ are related to each other in the range of $b=i\b$ for 
which they are both well defined. Let us stress that they cannot be 
the same, since their behavior under the two shifts differs by an 
exponential of an expression that is linear in $\a$. This suggest that 
the quotient of $\Y_b$ and $Y_\b$ can be expressed through Jacobi 
$\vartheta$-functions. In fact it is possible to show that in the 
region where  ${\rm Im} (b^2) = - {\rm Im} (\b^2) > 0$, the two 
functions are related by 
\begin{equation} \label{ZYYrel}
\Y_{i\b} (\a) \ = \ {\cal N}(\b)\, e^{-\frac{\pi i}{2}(Q_b/2 - \a)^2}
 \, e^{- \pi \frac{\a}{\b}}\, \vartheta_1(i\a\b^{-1},\b^{-2})
 \ Y_\b(\a) \ \ .       
\end{equation} 
Here, ${\cal N}(\beta)$ is some factor that will not be relevant for 
the following and $\vartheta_1$ is one of Jacobi's $\vartheta$-functions
(see appendix 2 for our conventions). One may test this formula simply 
by checking that both sides of the equality have the same behavior under 
the shifts $\a \ra \a + b^{\pm 1}$. Another direct proof is spelled
out in appendix 1.  
\smallskip

We are now prepared to spell out our proposal for the 3-point function 
of Liouville theory with $c \leq 1$. What we have shown so far is that 
\begin{eqnarray} \label{Cc<}   
 C(\a_1,\a_2,\a_3) & = & \left[\pi \mu \gamma(-\b^2) (-\b^2)^{1+\b^2}
\right]^{i (2\ta - Q)/\b} \ P(\a_1,\a_2,\a_3) \ e^{Q(\a_1,\a_2,\a_3)} 
\\[2mm] \label{P} 
\mbox{ where } & & P(\a_i) \ = \ e^{- 2 \pi i Q_b \ta} \, 
    \frac{{\cal N}_0(\b)\ \vartheta_1'(0)}{\vartheta_1(i(2\ta-Q_b)\b^{-1})}
  \ \prod_{j=1}^3 \, \frac{\vartheta_1(i2\a_j \b^{-1})}
                         {\vartheta_1(i2\ta_j \b^{-1})} 
\end{eqnarray}
and the function $Q$ was defined in eq.\ (\ref{Q}). In the derivation 
we used that $\vartheta_1(x) = \vartheta_1(x,\b^{-2})$ vanishes at 
$x=0$ and we absorbed a few $\beta$ dependent constants into the 
overall normalization ${\cal N}_0$. By construction, the factor
$\exp Q$ has a smooth continuation to the real $\beta$-axis. Our 
main claim is that (for appropriate choice of ${\cal N}_0(\beta)$) 
also $P$ remains well defined for real $\beta$, at least after 
restricting the labels $\a_j$ to some subset of the complex plane. 
In general, however, $P$ will not be an analytic function of the 
$\a_j$ but rather a distribution. We will compute this distribution 
for the $c=1$ theory below. 

A very naive inspection of our formula for $P$ could actually suggest 
that $P$ does not exist at $\beta^2 = 1$. In fact, as we take $\beta$ 
to the real axis, the modular parameter $q$ is sent to the boundary
of the unit disc and it is well known that Jacobi $\vartheta$-functions
are very singular for $|q| = 1$. At $\tau = \beta^{-2}=1$, for 
example, $\vartheta_1(x)$ is a periodic $\delta$-function in the real 
variable $x$ (see e.g.\ \cite{Mumford}). Since products and quotients of 
a $\delta$-function are not well defined, one might suspect that the 
same is true for $P$. Obviously, in this argument we have taken the 
limit too early. The correct prescription is to build $P$ from the 
$\vartheta$-functions first while $\b^2$ is still within the lower 
half-plane and then to take the limit of the resulting family of 
functions.   
\smallskip 

We conclude this section with a short argument showing that the new 
factor $P$ does resolve the mismatch between our earlier analysis 
of the minisuperspace limit of $\exp Q$  and the toy model from 
section 2. In fact, presenting $\vartheta_1$ as an infinite sum 
it is easy to see that
$$\lim_{\b \ra 0} q^{-\frac{1}{4}} \vartheta_1(2 i\omega,\beta^{-2}) 
    \ = \ 2 \sinh 2 \pi \omega \ \ .$$ 
Here, the point $\beta = 0$ is approached such that the imaginary 
part of $\beta^2$ is negative. Hence, $q = \exp i\pi /\b^2$ tends 
to zero and so only the leading term in $\vartheta_1$ can survive.
Using this simple result, we can calculate the minisuperspace
limit $\beta \ra 0$ with labels (\ref{aom}) inserted into the 
couplings (\ref{Cc<}). The result agrees exactly with the couplings
$C_0(\w_1,\w_2,\w_3)$ of the toy model provided that we set $\lambda
= 4 \pi \mu$.

\section{The $c=1$ Liouville theory}
\setcounter{equation}{0}
\def\te{\tilde \eta}
\def\vt{\vartheta} 

We will finally approach the main goal of this note: in this section 
we shall obtain two possible $c=1$ theories as a limit from Liouville
theory. The first one is defined for purely imaginary labels $\a$ and
we will identify it as the interacting $c=1$ theory that was constructed 
by Runkel and Watts \cite{Runkel:2001ng} as a limit of unitary minimal 
models. The second limit we take is defined for real $\a$ and it gives 
the 3-point couplings of the rolling tachyon solution. They agree with 
the proposal of Strominger and Takayanagi \cite{Strominger:2003fn} only 
for a certain subset of parameters $\a_j$. 
\medskip 

Let us first try to analyse the $c=1$ limit for $\a = i \eta$ purely 
imaginary. We claim that 
\begin{eqnarray}
P^\E_{c=1}(\eta_1,\eta_2,\eta_3) & = & \lim_{\beta \ra 1} e^{2\pi Q_b \te} \, 
    \frac{\b^{-1} Q_b\ \vartheta_1'(0)}{\vartheta_1((-2\te-iQ_b)\b^{-1})}
  \ \prod_{j=1}^3 \, \frac{\vartheta_1(-2\eta_j \b^{-1})}
                         {\vartheta_1(-2\te_j \b^{-1})} \\[2mm]
  & = & \frac{\pi}{2} e^{4\pi i \te} \left( \sum_{j=1}^3 \theta(\te_j) 
     - \theta(\te)\right) \left(1 + \theta(\te) \prod_{j=1}^3 
      \theta(\te_j)\right).    
\end{eqnarray} 
Here, $\theta$ is the periodic step function with period length $L=1$ 
(see appendix 2) and the limit is taken such that ${\rm Im}(\b) <0$. 
This formula is proved using a number of identities between products 
of theta functions, starting from the following equation (see e.g.\ 
\cite{Mumford})
\begin{eqnarray*} 
 &  &   \vt_1(2x)\, {\prod}_{j} \vt_1(-2\eta_j\b^{-1})+ 
  \vt_3(2x)  {\prod}_{j} \vt_3(-2\eta_j\b^{-1}) \\[2mm] 
 &  & \hspace*{1cm} \ = \   
  \vt_1(x-2\te\b^{-1} )  {\prod}_{j} \vt_1(x+2\te_j\b^{-1})
  +\vt_3(x-2\te\b^{-1})  {\prod}_{j} \vt_3(x+2\te_j\b^{-1})\ \ . 
\end{eqnarray*}  
We differentiate this with respect to $x$ and then set $x=0$. Since
$\vt_3'(0)=0$, the second term on the left hand side drops out and we 
can use the resulting equality to replace the product of $\vt$-functions
in the numerator of $P$ by a sum of two fourfold products with arguments
being the same as in the denominator of $P$. The evaluation of the limit 
is thereby reduced to the evaluation of the limit for $\vt_1'/\vt_1$, 
$\vt_3'/\vt_3$ and $\vt_3/\vt_1$. These limits are computed in appendix 2.
\smallskip

Our result demonstrates several of the general points we have made at the 
end of the previous section. Most importantly, the limit is completely 
well defined. But the resulting 3-point couplings are discontinuous in 
$\a = i \eta$ and hence there is no way to continue them naively beyond 
the real line. That is very different from the behavior of correlators 
at $c > 1$. 
\smallskip 

One may wonder whether we can really trust this outcome of our analysis. 
It could certainly happen that the limit is well defined but does not
give consistent crossing symmetric couplings of a conformal 
field theory. But in the present situation, there is very strong 
additional evidence for the existence of this $c=1$ model. In fact, 
some years ago, Runkel and Watts constructed an interacting 
non-rational $c=1$ theory as a limit of unitary minimal models
\cite{Runkel:2001ng}. Their 3-point couplings also split into two 
factors. One of them possesses a nice integral representation, while
the second jumps between $0$ and $1$. For this theory, Runkel and 
Watts tested the crossing symmetry numerically. We claim now that 
our couplings are related to those of the Runkel-Watts theory by 
a rather simple re-normalization of the fields.  

The agreement of the smooth factor $\exp Q$ in the Runkel-Watts 
solution with the factor $\exp Q$ in our theory is obvious 
from the discussion in appendix A.1 of \cite{Runkel:2001ng}. The 
relation between discontinuous factors can be seen if we rewrite 
our $P^\E$ in the following form 
\begin{equation} \label{PE2}            
P^\E_{c=1} \ = \ \pi  \,   \prod_{j=1}^3 
  e^{2\pi i \eta_j}\,  \theta(2\eta_j) \, \cdot\, \frac{1}{4}\, 
   \left(1 + \theta(\te) \prod_{j=1}^3  \theta(\te_j)\right).  
\end{equation} 
Except for the first factors that can be absorbed in a change 
of normalization of the fields, this is the same as the function $P$ 
defined in \cite{Runkel:2001ng}, though the detailed comparison is a 
bit messy since the authors of \cite{Runkel:2001ng} did not provide 
a closed expression for this function. 
\smallskip

We have emphasized at several places throughout this text how important 
poles in the 3-point function are to recover a sensible 2-point function
in the limit $\a_2 \ra 0$. But now we see that neither $\exp Q$ nor 
$P^E_{c=1}$ has poles at $\eta_1 - \eta_3 \pm \eta_2 =0$. Fortunately, 
the problem is rather easy to understand: when we formulated such a 
requirement on the existence of poles before, we assumed that the 
identity field is simply the limit of $\Phi_\a$ as $\a \ra 0$. But 
this need not be the case. In fact, Runkel and Watts explained already 
how to cure the issue by constructing the identity field as a limit of 
the derivative $\partial_\eta \Phi_{i\eta}$. We refer the reader to 
\cite{Runkel:2001ng} for details. The only new information that the 
2-point function contains is the reflection amplitude which we can 
read off directly from the 3-point couplings (see discussion at the 
end of section 3). Using the invariance of $Y_{\beta = 1}$ under 
reflection $\eta \ra -\eta$ we obtain $R^\E_{c=1}(\eta) \ = \ 
 - e^{2 \pi i \eta}$. Let us stress, however, that the result includes
a contribution from the factor that contains the coupling $\mu$. To hide 
the divergence of $\gamma(-\b^2)$ as $\beta \ra 1$, one has to introduce 
a renormalized coupling $\mu_{\rm ren}$. The conventions we have used 
here are the same as in \cite{Strominger:2003fn} and they are 
incorporated in our formula (\ref{Cint}).
\bigskip

Encouraged by the success of the Euclidean limit we now turn to a second 
$c=1$ limit that we take when the labels $\a$ are real. The analysis 
turns out to be a bit simpler than in the previous case. Using the 
modular properties of $\vt_1$ we can show 
\begin{eqnarray} 
P^\L_{c=1}(\a_1,\a_2,\a_3) & = & \lim_{\b \ra 1}  e^{-2\pi i Q_b \ta} \, 
    \frac{\b^{-1} Q_b \ \vartheta_1'(0)}
          {\vartheta_1(i(2\ta-Q_b)\b^{-1})}
  \ \prod_{j=1}^3 \, \frac{\vartheta_1(i2\a_j \b^{-1})}
                         {\vartheta_1(i2\ta_j \b^{-1})} \\[2mm] 
& = &  e^{4\pi \ta}\,  \lim_{\epsilon \ra 0^+}   
  \frac{\pi}{\sin\pi (\frac{\ta}{\e}+ i\ta)}\ \prod_{j=1}^3 \ 
   \frac{\sin\pi (\frac{\a_j}{\e}+i\a_j)}
     {\sin \pi (\frac{\ta_j}{\e}+i\ta_j)}
\end{eqnarray} 
for $\a_j \in \QR^+$. To compute the limit, we would like to rewrite 
the $\sin$-factors in the denominator through a geometric series 
expansion. This requires to distinguish between four different 
regions, depending on the signs of $\ta_j$. Since $\a_j$ are positive, 
the same is true for $\ta$. But the signs of $\ta_j$ vary. They can 
either be all positive, or it can happen that one of the $\ta_j$ is 
negative. In each of these cases, an expansion can be performed and 
we are left with an infinite sum of exponentials. When we take  the 
limit $\e \ra 0$, terms containing an $\exp i \a_j/\e$ may be neglected
since the 3-point function is considered as a distribution. Hence, we 
only have to look for the constants in our expansion. But 
such constant terms appear exclusively in the region where all $\ta_j > 0$. 
This result can be spelled out more formally using the step function 
\begin{equation}\label{Theta} 
\Theta(x) \ := \ \, \left\{ \begin{array}{ll} 
   -1 \ \ \ \ & \mbox{ for } \ \ \ \ \   x < 0 \\[1mm]
   +1        & \mbox{ for } \ \ \ \ \  0 < x 
  \end{array} \right. \ \ . 
\end{equation}        
From our above discussion we infer that for $\a_j > 0$  
\begin{equation} 
P^\L_{c=1}(\a_1,\a_2,\a_3) \ = \ \pi i e^{2\pi \sum_j \a_j} \left( 1 + 
    \Theta(\ta) \ \prod_{j=1}^3 \, \Theta(\ta_j) \right) \ \ . 
\end{equation} 
Note that the form of the final expression for $P^\L_{c=1}$ is very similar 
to the corresponding formula (\ref{PE2}) for its Euclidean counterpart, only 
that the periodic step function $\theta$ has been replaced by $\Theta$. In 
the region where all the $\ta_j$ are positive, our result for $P^\L$ agrees 
with the findings of \cite{Strominger:2003fn}. But the appearance of the 
step functions resolves a puzzle concerning the 2-point function of the 
model: As observed in \cite{Strominger:2003fn}, their formula for the 
3-point function did not reproduce the right 2-point function upon 
sending $\a_2 \ra 0$. In our result, the issue can be settled in the 
same way as it was settled for the Runkel-Watts theory. Once more, the 
reflection amplitude can be found without going into the details of 
how to construct the identity field. Analysing the behavior of 
$C^\L_{c=1}$ under reflections $\a_1 \ra -\a_1$ we find $R^\L_{c=1} (\a)=  
\exp 2\pi a$. As for the Euclidean theory, this result depends on how 
the subtleties in the definition of $\mu_{\rm ren}$ are dealt with. Here 
we have followed the conventions of \cite{Strominger:2003fn}.

\section{Conclusion and outlook} 
\setcounter{equation}{0}

In this note we have proposed an exact solution for Liouville theory 
with $c\leq 1$. The formulas were obtained through a rigorous limit
from the solution of the model at $c \geq 25$. Explicit expressions 
have been worked out for $c=1$. In this case we found two different 
limiting theories with real conformal weights, depending on the range 
of the labels $\a$. The Euclidean limit coincides with an interacting 
$c=1$ theory that was first constructed by Runkel and Watts as a limit 
of unitary minimal models. Our formulas for the Lorentzian limit, 
however, are new and provide a very promising starting point for 
further studies of tachyon decays in string theory. 
\smallskip 

Even though understanding the homogeneous decay of bulk tachyons is 
certainly an important problem in (bosonic) string theory, the 
corresponding processes are expected to be rather violent. The 
condensation of open string tachyons are somewhat better behaved, 
mostly because such processes can be probed with closed strings. 
Finding expressions for 2- and 3-point functions of boundary 
Liouville theory, including the 2-point function between one bulk 
and one boundary field, is therefore of considerable interest. We 
have claimed in the introduction that the solution of the bulk 
problem presents an important step toward solving the boundary 
theory (with and without bulk coupling). Now we can see precisely
how far we got. In fact, our analysis here was centered around 
understanding the function $\Y$ which is built from a product 
of two double $\Gamma$-functions. But for the solution of the 
boundary theory we need to investigate one half of $\Y$, i.e.\ Barnes' 
double $\Gamma$-function itself (see e.g.\ \cite{Fateev:2000ik,
Teschner:2000md,Ponsot:2001ng,Hosomichi:2001xc,Zamolodchikov:2001ah} 
for some formulas concerning the models with $c \geq 25$). We shall 
address the relevant issues in a forthcoming publication 
\cite{FreSch}. 
Note that in the case of open strings, even the 2-point function 
is built from Barnes' double $\Gamma$-function and hence its 
$c=1$-limit is expected to be quite non-trivial. A formula for
this 2-point function was proposed recently in \cite{Gutperle:2003xf}, 
but with our new techniques at hand it would be reassuring to confirm this 
proposal through a more rigorous derivation. Another interesting 
quantity that has not been studied yet is the 2-point function of
a bulk- and a boundary field.  
\smallskip

Let us finally stress that our analysis should also carry over to 
several other interesting decay processes. Most importantly, using 
the Sine-Liouville model (see e.g.\ \cite{Fukuda:2002bv} for an exact 
solution) it is possible to study the decay of a bulk tachyon with a 
$\sin$-shaped profile in one space-like direction $X$. In the case of 
boundary perturbations, such profiles have received a lot of attention 
(see e.g.\ \cite{Callan:1994ub,Polchinski:1994my,Fendley:1994rh,
Recknagel:1998ih, Gaberdiel:2001zq} for boundary conformal field theory 
treatments) 
because they allow to interpolate between Neumann and Dirichlet 
boundary conditions. One attractive feature of such models is that 
they possess a tunable parameter, namely the period length $L=1/a$ 
of the profile. This allows to bring the relevant field $\Phi = \sin
2\pi a X$ arbitrarily close to marginality. It would be interesting to 
compare exact solutions of both the bulk and the boundary Sine-Liouville
model for $c=2$ with RG studies in the Sine-Gordon model, especially in 
the regime where the tachyon profile becomes marginal. We leave these
problems to future investigations.        
\bigskip
\bigskip 
\bigskip 

\noindent
{\bf Acknowledgement:} I am grateful to J.\ de Boer, S.\ Fredenhagen,
I.\ Kostov, B.\ Ponsot, A.\ Recknagel, S.\ Ribault, I.\ Runkel and Al.B.\ 
Zamolodchikov 
for interesting comments and discussions. It is a special pleasure to thank 
J.\ Teschner for numerous conversations in which he shared many of his 
insights into Liouville theory. My interest in this problem was particularly 
stimulated through the interaction with M.\ Gutperle and A.\ Strominger. 
\newpage

\section{Appendix 1: Barnes' $\Gamma$-function, $\Y$ and $Y$}
\setcounter{equation}{0}

We use this first appendix to collect some definitions and standard 
results concerning Barnes' double $\Gamma$-function (see also e.g.\ 
\cite{Barnes1,Jimbo:1996ss}). These are used 
throughout the main text. We then spell out a direct proof of our 
relation (\ref{ZYYrel}). 
\smallskip 

Barnes' double $\Gamma$-function $\Gamma_a(y) = \Gamma_2(y|a,a^{-1})$ is 
defined for $y \in \QC$ and complex $a$ with ${\rm Re}(a) \neq 0$ such 
that its logarithm possesses the following integral representation, 
\begin{equation}\label{BGamma}
 \ln \Gamma_a(y) \ = \ c_a + \int_0^\infty \frac{d\t}{\t}
  \left[ \frac{e^{-y\t} - e^{- Q_a \t/2}} 
              {(1-e^{-a \t}) (1- e^{-\t/a})} 
         - \frac{\left(\frac{Q_a}{2} - y\right)^2}{2} e^{-\t}
           - \frac{\frac{Q_a}{2} -y }{\t}\right]  
\end{equation}
where $Q_a = a + a^{-1}$ and $c_a = \ln \Gamma_2(Q_a/2|a,a^{-1})$ is a 
constant. The integral exists for $0 < {\rm Re}(y)$. Under shifts 
by $a^{\pm 1}$ the function $\Gamma_a$ behaves according to 
\begin{equation}\label{BGshift}
\Gamma_a(y+a) \ = \  \sqrt{2\pi} \, \frac{a^{ay-\frac12}}{\Gamma(ay)}\, 
                      \Gamma_a(y) \ \ , \ \ 
\Gamma_a(y+a^{-1}) \ = \ \sqrt{2\pi} \, \frac{a^{-\frac{y}{a}+\frac12}}
      {\Gamma(a^{-1}y)}\, \Gamma_a(y) \ \ . 
\end{equation}
The functions $\Y_b$ and $Y_\b$ defined through eqs.\ (\ref{ZY}) 
and (\ref{Y}) in the main text are both constructed out of products 
of Barnes double $\Gamma$-function and it is easy to derive some of their 
main properties using the above formulas for $\Gamma_a$. 

To prove the formula (\ref{ZYYrel}) we depart from the integral 
representation (\ref{ZYpsilon}) and rotate the contour onto the 
positive imaginary axis, 
\begin{eqnarray*} 
  \ln \Y_b(\a) & = &  -2c_b +  \int_0^\infty  \frac{d\t}{\t}
              \left[ e^{-i\t}\left(\frac{Q_b}{2}-\a\right)^2   -
               \frac{ \sin^2\left(\frac{Q_b}{2}-\a\right)\frac{\t}{2}}
               {\sinh\frac{\b \t}{2} \sinh\frac{\t}{2\b}}    
               \right]    \\[2mm] 
   &  & \hspace*{3.5cm} 
     + 2i \sum_{n=1}^\infty \frac{(-1)^n}{n} 
   \frac{\sin^2 \frac{\pi i n}{\b}(Q_b/2 -\a)}{\sin \frac{\pi n}{\beta^2}}
\end{eqnarray*} 
where $\b = i \beta$, as usual. The second term is the contribution 
from the poles that we cross while 
rotating the contour. Note that, according to the integral formula 
(\ref{Yint}), the first term can be expressed easily through our 
function $Y_\b$. In addition, we rewrite the infinite sum by means 
of some simple trigonometric identities, 
$$ 
  \ln \Y_b(\a) \, \sim\,  \ln Y_\b (\a) - \frac{\pi i}{2} \, 
   \left(\frac{Q_b}{2}-\a\right)^2 
 -i \sum_{n=1}^{\infty} \frac{1}{n} \, 
     \sin \frac{2\pi  n i \a }{\b} + \frac{1}{n}\,   
     \cot\frac{\pi n}{\b^2} \, \cos \frac{2\pi n i \a}{\b}
$$ 
up to some constant depending only on $b$. It will drop out 
once we combine the $\Y$-function into a formula for the 3-point 
coupling. Our next step is to expand the factor $\cot\frac{\pi n}{\b^2}$. 
In the region where ${\rm Im}(\b^2) < 0$, we have    
$$   \cot\frac{\pi n}{\b^2} \ = \ -i \left ( 1 + 2 \sum_{\nu = 1}^{\infty} 
              e^{\frac{2\pi i n \nu}{\b^2}}\right) \ \ . $$ 
We can now carry out the summation over $n$ in the previous formula
for $\ln\Y$ to find that 
$$  
 \ln \Y_b(\a) \, \sim \,  \ln Y_\b (\a) - \frac{\pi i}{2} \, 
   \left(\frac{Q_b}{2}-\a\right)^2 
+ \ln (1-e^\frac{2\pi i \eta}{\b})\prod_{\nu =1}^\infty 
  (1-q^{2\nu} e^\frac{2\pi i \eta}{\beta})(1-q^{2\nu} 
e^\frac{-2\pi i \eta}{\b}) 
$$ 
where $q = \exp (\pi i/\b^2)$. In the last term we recognize the product 
representation of a $\vartheta$-function (see appendix 2) and hence 
we arrive at the formula (\ref{ZYYrel}).

\section{Appendix 2: Formulas for the $c=1$ limit} 
\setcounter{equation}{0}
\def\e{\epsilon} 
\def\vt{\vartheta}

In this technical appendix we state and prove several facts concerning 
the limit of various combinations of $\vt$-functions as the modular 
parameter $\tau = 1/\beta^2$ is sent to the real line. Most of the 
properties of $\vartheta$-functions that we will use can be found e.g.\ 
in \cite{Mumford}. Let us begin with the definition of the functions 
$\vt_1$ and $\theta$. In our conventions, the former is given by 
\bea \vartheta_1(x,\tau) & := & - 2 q^{\frac{1}{4}} \sin \pi x 
  \ \prod_{n=1}^{\infty}\, (1-q^{2n})(1-q^{2n} e^{2\pi i x}) 
   (1-q^{2n}e^{-2\pi i x}) \\[2mm] 
 & = & \sum_{m= -\infty}^\infty \, q^{(m+\frac12)^2} \, 
       e^{2\pi i (m+\frac12)(x+\frac12)} \ \  
\eea     
where $q = \exp \pi i \tau$. We also use the Jacobi $\vt$-function 
$\vt_3$ which may be defined through the following sum
$$ \vt_3(x,\t) \ = \  \sum_{m= -\infty}^\infty \, q^{m^2} \, 
       e^{2\pi i m x}\ \ . 
$$ 
Finally, we introduce a periodic function $\theta(x)= \theta(x+1)$ 
by 
\begin{equation}\label{theta} 
\theta(x) \ := \ \, \left\{ \begin{array}{ll} 
   -1 \ \ \ \ & \mbox{ for } \ -1/2 < x < 0 \\[1mm]
   +1        & \mbox{ for } \ \ \ \ \  0 < x < 1/2 
  \end{array} \right. \ \ . 
\end{equation} 
In our evaluation of the $c=1$ limit of the 3-point function of 
Liouville theory we employ three simple formulas that we spell 
out now. The first of these formulas is  
\begin{equation}\label{lemma1} 
\lim_{\e\ra 0^+} \, \e\, \frac{\vt'_1(x,1+i\e)}{\vt_1(x,1+i\e)} \ = \ 
       - 2\pi x + \pi  \theta(\frac12x)  \ \ \mbox{ for all } \ \ x 
   \in \, (-1,1) \ \ . 
\end{equation}
To compute the limit on the left hand side, we use the behavior of 
$\vt_1$ under the modular transformations $\tau \ra \tau - 1$ and $\tau 
\ra - 1/\tau$ and obtain
\beas  
\lim_{\e\ra 0^+} \, \e\, \frac{\vt'_1(x,1+i\e)}{\vt_1(x,1+i\e)} 
& = & \lim_{\e\ra 0^+} \, \e\, \frac{d}{dx} \ln 
     \left( - \e^{-\frac12} e^{-\pi \frac{x^2}{\e}} 
             \vt_1(\frac{x}{i\e},\frac{i}{\e})\right) \\[2mm] 
& = & -2\pi x - i \lim_{\e\ra 0^+} \frac{\vt_1'(\frac{x}{i\e},\frac{i}{\e})}
               {\vt_1(\frac{x}{i\e},\frac{i}{\e})} \ \ . 
\eeas 
The second term is then evaluated by means of the expansion (see e.g.\ 
\cite{Abramowitz})  
$$ \frac{\vt_1'(y,\t)}{\vt_1(y,\t)} \ = \ 
     \pi \cot \pi y + 4 \pi \sum_{n=1}^\infty \, 
           \frac{q^{2n}}{1-q^{2n}} \sin 2\pi n y \ \  
$$  
which we can use in the region $|{\rm Im}(y)| < {\rm Im}(\t)$ where 
the sum is absolutely convergent.  
This concludes the proof of eq.\ (\ref{lemma1}). Our second formula  
\begin{equation}\label{lemma2}
\lim_{\e\ra 0^+} \, \e\, \frac{\vt'_3(x,1+i\e)}{\vt_3(x,1+i\e)} \ = \ 
       - 2\pi x + \pi \theta(\frac12 x) \ \ \mbox{ for all } 
   \ \ x \in \, (-1,1) \ \ . 
\end{equation}
is very similar to the first one and it is proved along the same 
lines. We leave this as an exercise. Finally, we also claim that 
\begin{equation}\label{lemma3}
\lim_{\e\ra 0^+} \, \frac{\vt_3(x,1+i\e)}{\vt_1(x,1+i\e)} \ = \ 
        i e^{- \frac{\pi i}{4}}\, \theta(\frac12 x) \ \ 
 \mbox{ for all } \ \  x \in \QR \ \ . 
\end{equation}
Modular properties of $\vt$-functions are employed in a first step to 
see that 
$$\lim_{\e\ra 0^+} \, \frac{\vt_3(x,1+i\e)}{\vt_1(x,1+i\e)}
\ = \ e^{-\frac{\pi i}{4}}\, \lim_{\e\ra 0^+} \, 
   \frac{\vt_1(\frac{x}{i\e}+\frac12,\frac{i}{\e})}
          {\vt_1(\frac{x}{i\e},\frac{i}{\e})} \ \ . $$ 
The limit is then evaluated with the help of the expansion (see e.g.\ 
\cite{Whittaker})  
$$  \ln \frac{\vt_1(y+\frac12,\t)}{\vt_1(y,\t)} \ = \ 
      \ln \cot \pi y + 4 \sum_{m=1}^\infty \, \frac{1}{2m-1} 
    \, \frac{q^{4m-2}}{1-q^{4m-2}} \, \cos \pi(4m-2)y  
$$ 
which holds for $|{\rm Im}(y)| < {\rm Im}(\t)$. 
The eqs.\  (\ref{lemma1}-\ref{lemma3}) provide the main ingredients in our 
computation of the Liouville 3-point function at $c=1$. This is explained 
in detail in Section 5.
\newpage


\begin{thebibliography}{99}


\baselineskip=18pt 

\bibitem{Martinec:2002tz}
E.~J. Martinec, {\it Defects, decay, and dissipated states},
{\tt hep-th/0210231}.

\bibitem{Gutperle:2002ki}
M.~Gutperle, M.~Headrick, S.~Minwalla and V.~Schomerus, {\it Space-time energy
  decreases under world-sheet {RG} flow},  {\em JHEP} {\bf 01} (2003) 073
  [{\tt hep-th/0211063}].

\bibitem{Sen:2002nu}
A.~Sen, {\it Rolling tachyon},  {\em JHEP} {\bf 04} (2002) 048
  [{{\tt hep-th/0203211}}].

\bibitem{Sen:2002vv}
A.~Sen, {\it Time evolution in open string theory},  {\em JHEP} {\bf 10} (2002)
  003 [{{\tt hep-th/0207105}}].

\bibitem{Strominger:2002pc}
A.~Strominger, {\it Open string creation by {S-branes}},
 {{\tt hep-th/0209090}}.

\bibitem{Gutperle:2003xf}
M.~Gutperle, A.~Strominger, {\it Timelike boundary {Liouville} theory},
 {{\tt hep-th/0301038}}.

\bibitem{Strominger:2003fn}
A.~Strominger and T.~Takayanagi, {\it Correlators in timelike bulk {Liouville}
  theory}, {{\tt hep-th/0303221}}.

\bibitem{Mukhopadhyay:2002en}
P.~Mukhopadhyay and A.~Sen, {\it Decay of unstable {D-branes} with electric
  field},  {\em JHEP} {\bf 11} (2002) 047
[{{\tt hep-th/0208142}}].

\bibitem{Okuda:2002yd}
T.~Okuda and S.~Sugimoto, {\it Coupling of rolling tachyon to closed strings},
  {\em Nucl. Phys.} {\bf B647} (2002) 101--116
  [{{\tt hep-th/0208196}}].

\bibitem{Chen:2002fp}
B.~Chen, M.~Li and F.-L. Lin, {\it Gravitational radiation of rolling tachyon},
   {\em JHEP} {\bf 11} (2002) 050
 [{{\tt hep-th/0209222}}].

\bibitem{Larsen:2002wc}
F.~Larsen, A.~Naqvi and S.~Terashima, {\it Rolling tachyons and decaying
  branes},  {\em JHEP} {\bf 02} (2003) 039
  [{{\tt hep-th/0212248}}].

\bibitem{Rey:2003xs}
S.-J. Rey and S.~Sugimoto, {\it Rolling tachyon with electric and magnetic
  fields: T- duality approach},  {\em Phys. Rev.} {\bf D67} (2003) 086008
  [{{\tt hep-th/0301049}}].

\bibitem{Lambert:2003zr}
N.~Lambert, H.~Liu and J.~Maldacena, {\it Closed strings from decaying
  {D-branes}}, {{\tt
  hep-th/0303139}}.

\bibitem{McGreevy:2003kb}
J.~McGreevy and H.~Verlinde, {\it Strings from tachyons: {The} c = 1 matrix
  reloated}, {{\tt hep-th/0304224}}.

\bibitem{Martinec:2003ka}
E.~J. Martinec, {\it The annular report on non-critical string theory}, 
{{\tt hep-th/0305148}}.

\bibitem{Klebanov:2003km}
I.~R. Klebanov, J.~Maldacena and N.~Seiberg, {\it D-brane decay in
  two-dimensional string theory},
 {{\tt hep-th/0305159}}.

\bibitem{McGreevy:2003ep}
J.~McGreevy, J.~Teschner and H.~Verlinde, {\it Classical and quantum {D-branes
  in 2D} string theory}, {{\tt
  hep-th/0305194}}.

\bibitem{Dorn:1994xn}
H.~Dorn and H.~J. Otto, {\it Two and three point functions in {Liouville}
  theory},  {\em Nucl. Phys.} {\bf B429} (1994) 375--388
  [{{\tt hep-th/9403141}}].

\bibitem{Zamolodchikov:1996aa}
A.~B. Zamolodchikov and A.~B. Zamolodchikov, {\it Structure constants and
  conformal bootstrap in {Liouville} field theory},  {\em Nucl. Phys.} {\bf
  B477} (1996) 577--605. 

\bibitem{Ponsot:1999uf}
B.~Ponsot and J.~Teschner, {\it Liouville bootstrap via harmonic analysis on a
  noncompact quantum group},  {{\tt
  hep-th/9911110}}.

\bibitem{Teschner:2001rv}
J.~Teschner, {\it Liouville theory revisited},  {\em Class. Quant. Grav.} {\bf
  18} (2001) R153--R222 [{{\tt
  hep-th/0104158}}].

\bibitem{Teschner:2003en}
J.~Teschner, {\it A lecture on the {Liouville} vertex operators},
 {{\tt hep-th/0303150}}.

\bibitem{Gaiotto:2003rm}
D.~Gaiotto, N.~Itzhaki and L.~Rastelli, {\it Closed strings as imaginary
  {D-branes}}, {{\tt
  hep-th/0304192}}.

\bibitem{Runkel:2001ng}
I.~Runkel and G.~M.~T. Watts, {\it A non-rational {CFT} with c = 1 as a limit
  of minimal models},  {\em JHEP} {\bf 09} (2001) 006
  [{{\tt hep-th/0107118}}].

\bibitem{Goulian:1991qr}
M.~Goulian and M.~Li, {\it Correlation functions in {Liouville} theory},  {\em
  Phys. Rev. Lett.} {\bf 66} (1991) 2051--2055.

\bibitem{Teschner:1995yf}
J.~Teschner, {\it On the {Liouville} three point function},  {\em Phys. Lett.}
  {\bf B363} (1995) 65--70 [{{\tt
  hep-th/9507109}}].

\bibitem{Mumford}
D.~Mumford, {\em {Tata lectures on theta functions}}.
\newblock Birkhaeuser, Boston, 1983.

\bibitem{Fateev:2000ik}
V.~Fateev, A.~B. Zamolodchikov and A.~B. Zamolodchikov, {\it Boundary
  {Liouville} field theory. {I: Boundary} state and boundary two-point
  function},  {{\tt hep-th/0001012}}.

\bibitem{Teschner:2000md}
J.~Teschner, {\it Remarks on {Liouville} theory with boundary},
  {{\tt hep-th/0009138}}.

\bibitem{Ponsot:2001ng}
B.~Ponsot and J.~Teschner, {\it Boundary {Liouville field theory: Boundary}
  three point function},  {\em Nucl. Phys.} {\bf B622} (2002) 309--327
  [{{\tt hep-th/0110244}}].

\bibitem{Hosomichi:2001xc}
K.~Hosomichi, {\it Bulk-boundary propagator in {Liouville} theory on a disc},
  {\em JHEP} {\bf 11} (2001) 044
  [{{\tt hep-th/0108093}}].

\bibitem{Zamolodchikov:2001ah}
A.~B. Zamolodchikov and A.~B. Zamolodchikov, {\it Liouville field theory on a
  pseudosphere},  {{\tt
  hep-th/0101152}}.

\bibitem{FreSch}
S.~Fredenhagen and V.~Schomerus, {\it {\it work in progress}}.

\bibitem{Fukuda:2002bv}
T.~Fukuda and K.~Hosomichi, {\it Super {Liouville} theory with boundary},  {\em
  Nucl. Phys.} {\bf B635} (2002) 215--254
  [{{\tt hep-th/0202032}}].

\bibitem{Callan:1994ub}
J.~Callan, Curtis~G., I.~R. Klebanov, A.~W.~W. Ludwig and J.~M. Maldacena, {\it
  Exact solution of a boundary conformal field theory},  {\em Nucl. Phys.} {\bf
  B422} (1994) 417--448 [{{\tt
  hep-th/9402113}}].

\bibitem{Polchinski:1994my}
J.~Polchinski and L.~Thorlacius, {\it Free fermion representation of a boundary
  conformal field theory},  {\em Phys. Rev.} {\bf D50} (1994) 622--626
  [{{\tt hep-th/9404008}}].

\bibitem{Fendley:1994rh}
P.~Fendley, H.~Saleur and N.~P. Warner, {\it Exact solution of a massless
  scalar field with a relevant boundary interaction},  {\em Nucl. Phys.} {\bf
  B430} (1994) 577--596 [{{\tt
  hep-th/9406125}}].

\bibitem{Recknagel:1998ih}
A.~Recknagel and V.~Schomerus, {\it Boundary deformation theory and moduli
  spaces of {D-branes}},  {\em Nucl. Phys.} {\bf B545} (1999) 233--282
  [{{\tt hep-th/9811237}}].

\bibitem{Gaberdiel:2001zq}
M.~R. Gaberdiel and A.~Recknagel, {\it Conformal boundary states for free
  bosons and fermions},  {\em JHEP} {\bf 11} (2001) 016
  [{{\tt hep-th/0108238}}].

\bibitem{Barnes1}
E.~Barnes, {\it The theory of the double gamma function},  {\em Philos. Trans.
  Roy. Soc.} {\bf A196} (1901) 265.

\bibitem{Jimbo:1996ss}
M.~Jimbo and T.~Miwa, {\it {QKZ} equation with $|q|=1$ and correlation
  functions of the {XXZ} model in the gapless regime},  {\em J. Phys.} {\bf
  A29} (1996) 2923--2958 [{{\tt
  hep-th/9601135}}].

\bibitem{Abramowitz}
M.~Abramowitz and I.~A. Stegun, {\em { Handbook of Mathematical Functions with
  Formulas, Graphs, and Mathematical Tables}}.
\newblock Dover, New York, 1972.

\bibitem{Whittaker}
E.~T. Whittaker and G.~N. Watson, {\em {A course in modern analysis}}.
\newblock Cambridge University Press, Cambridge, 1990.

\end{thebibliography}
\end{document}